\documentclass[11pt, a4paper]{article}
\usepackage{amssymb}
\usepackage{epsfig,sint,macros,cite,mathbbol}
\begin{document}
\begin{titlepage}
\begin{flushright}
\hfill CERN-PH-TH/2008-040\\
\hfill IFIC/08-13       \\
\hfill IFT-UAM-CSIC-08-13\\
\hfill FTUAM-08-05 \\
\hfill Edinburgh 2008/10\\
\hfill MKPH-T-08-03\\
\end{flushright}

\vskip 0.5 cm
\begin{center}
  {\Large\bf Testing chiral effective theory with quenched lattice QCD \\[0.5ex] } 
\end{center}
\vskip 0.5 cm
\begin{center}
{\large 
     L. Giusti$^{\scriptscriptstyle a,1}$, P. Hern\'andez$^{\scriptscriptstyle b,2}$, S. Necco$^{\scriptscriptstyle b,3}$,
C. Pena$^{\scriptscriptstyle c,4}$, \\J. Wennekers$^{\scriptscriptstyle d,5}$, H. Wittig$^{\scriptscriptstyle e,6}$. 
}
\vskip 0.5cm
$^{\scriptstyle a}$
CERN, Physics Department, TH Division, CH-1211 Geneva 23, Switzerland
\vskip 1.5ex
$^{\scriptstyle b}$
Instituto de F\'isica Corpuscular, CSIC-Universitat de Val\`encia\\
Apartado de Correos 22085, E-46071 Valencia, Spain
\vskip 1.5ex
$^{\scriptstyle c}$
Dpto. F\'isica Te\'orica and Instituto de F\'isica Te\'orica UAM/CSIC\\
Facultad de Ciencias, Universidad Aut\'onoma de Madrid\\
Cantoblanco, E-28049 Madrid, Spain
\vskip 1.5ex
$^{\scriptstyle d}$
School of Physics, University of Edinburgh, Edinburgh EH9 3JZ, UK
\vskip 1.5ex
$^{\scriptstyle e}$
Institut f\"ur Kernphysik, University of Mainz, D-55099 Mainz, Germany
\vskip 1.0cm
{\bf Abstract}
\vskip 0.35ex
\end{center}
\noindent
We investigate two-point correlation functions of left-handed currents computed in quenched lattice QCD with the Neuberger-Dirac operator. 
We consider two lattice spacings $a\simeq 0.09,0.12$ fm and two different lattice extents $L\simeq 1.5, 2.0$ fm; quark masses span both the $p$- and the $\epsilon$-regimes. We compare the results with the predictions of quenched chiral perturbation theory, with the purpose of testing to what extent the effective theory reproduces quenched QCD at low energy.
In the $p$-regime we test volume and quark mass dependence of the pseudoscalar decay constant and mass; in the $\epsilon$-regime, we investigate volume and topology dependence of the correlators. 
While the leading order behaviour predicted by the effective theory is very well reproduced by the lattice data in the range of parameters that we explored, our numerical data are not precise enough to test next-to-leading order effects.

\noindent

\vskip 1.0 cm

\noindent
{\footnotesize $^{\scriptstyle 1}$ Leonardo.Giusti@cern.ch}, 
{\footnotesize $^{\scriptstyle 2}$ pilar.hernandez@ific.uv.es}, 
{\footnotesize $^{\scriptstyle 3}$ silvia.necco@ific.uv.es},
{\footnotesize $^{\scriptstyle 4}$ carlos.pena@uam.es},
{\footnotesize $^{\scriptstyle 5}$ jwenneke@ph.ed.ac.uk},
{\footnotesize $^{\scriptstyle 6}$ wittig@kph.uni-mainz.de}.

\vfill

\eject

\vfill

\eject

\end{titlepage}

\section{Introduction}
Lattice QCD underwent continuous advances in recent years, thanks to increasing computing resources and to very important theoretical and algorithmic improvements.
One of the most important outcomes of these efforts is that unquenched computations are now reaching ranges of volumes and quark masses where the QCD chiral dynamics can be tested and the matching with the chiral effective theory be performed.
Several studies for $N_{\rm f}=2$ have been recently undertaken, with Wilson fermions (plain and with $O(a)$ improvement \cite{DelDebbio:2006cn,DelDebbio:2007pz}) and Wilson twisted mass fermions \cite{Boucaud:2007uk,Blossier:2007vv,Boucaud:2008xu}. 
For the case $N_{\rm f}=3$, beside the results published in 2004 obtained with staggered fermions \cite{Aubin:2004fs} (see \cite{Bernard:2007ps} for
recent updates), new data are available for Domain Wall fermions \cite{Allton:2007hx} and for Wilson fermions \cite{Ukita:2007cu}. 
Simulations with pseudoscalar masses $M_P$ as low as $200-300$~MeV are
becoming state-of-the-art. Therefore, it is highly important to
understand to what extent the chiral effective theory can be applied
in the region, in order to perform controlled chiral extrapolations.\\
On the other hand, in the past years many studies have matched quenched lattice results with the chiral effective theory \cite{Heitger:2000ay,Aoki:2002fd,Bardeen:2003qz,Giusti:2004yp,Babich:2005ay}, using the non-trivial assumption that quenched QCD at low energy is described by quenched chiral effective theory. 
The main goal of this work is to verify this assumption in two different kinematic corners of the chiral regime of QCD.

On a finite volume with linear extent $L$, apart from the conventional $p$-regime, where finite-volume effects are exponentially suppressed in $(M_PL)$, one can extract information from other kinematic regions, for instance the $\epsilon$-regime, where the chiral limit $m\rightarrow 0$ is taken while keeping the pion wavelength much larger than $L$ \cite{Gasser:1987ah,Gasser:1986vb}. Higher order corrections in the chiral theory for physical observables look very different in the two cases;  
the matching between QCD and the chiral effective theory gives rise to different systematic effects and allows to extract quantities in the chiral limit from different observables. On the lattice, several quenched simulations in the $\epsilon$- regime with Ginsparg-Wilson fermions have been performed \cite{Hernandez:1999cu,DeGrand:2001ie,Hasenfratz:2002rp,Bietenholz:2003bj,Giusti:2003iq,Giusti:2004yp,Fukaya:2005yg,Bietenholz:2006fj,Giusti:2007cn}. In the quenched approximation it is now possible to collect results with Ginsparg-Wilson fermions at different volumes and lattice spacings with high statistics and reasonable computational effort. On the other hand, in full QCD the simulation of the Ginsparg-Wilson Dirac operator is still very expensive from the numerical point of view. First results for $N_{\rm f}=2$ are available both in the $p$- and $\epsilon$-regime \cite{Lang:2006ab,Fukaya:2007yv,Fukaya:2007pn,Aoki:2008ss,Noaki:2007es,Hasenfratz:2007yj,DeGrand:2006nv,DeGrand:2007tm}, although at the moment only for limited ranges of lattice spacings and volumes. For other regularisations, which break explicitly chiral symmetry at finite lattice spacing, exploring the $\epsilon$-regime is problematic.

We will adopt the Neuberger solution of the Ginsparg-Wilson relation for the Dirac operator.
In particular we will compute the left-handed current two-point correlation function both in the $p$- and in the $\epsilon$-regimes; in the $p$-regime we will investigate the quark mass and volume dependence of the pseudoscalar decay constant and mass, and compare it with the expectations from the chiral effective theory at NLO.
In the $\epsilon$-regime, we will also study the topology dependence of the current correlators.
We shall extract the corresponding Low Energy Couplings (LECs) and compare the results obtained for the leading order constants $\Sigma$ and $F$ in the two regimes. Moreover, by considering different lattice spacings, we will check that lattice artifacts are small, as already observed in many computations using the Neuberger Dirac operator (see e.g \cite{Wennekers:2005wa}). This work complements and expands the study published in \cite{Giusti:2004yp,Wennekers:2005wa}.\\
The paper is organised as follows: in Section \ref{sec2} we recall the main results from the quenched chiral effective theory at NLO, in the $\epsilon$- and $p$-regime; Section \ref{sec3} is devoted to describing the details of our numerical simulations; in Sections \ref{sec4} and \ref{sec5} we match the lattice results with the chiral effective theory in the $\epsilon$- and $p$-regimes respectively; finally, in Section \ref{sec6} we compare and discuss the results obtained for $F$ and $\Sigma$ in the two cases.

\section{Current correlator in the chiral effective theory}\label{sec2}
At leading order, the Euclidean Lagrangian of the chiral effective theory is given by \cite{Weinberg:1978kz,Gasser:1983yg}
\begin{equation}
\mathcal L = \frac{F^2}{4}{\rm Tr}\left\{\partial_{\mu} U^\dagger \partial_{\mu}U           \right\}-\frac{\Sigma}{2}{\rm Tr} \left\{e^{i\theta/N_{\rm f}}U\mathcal{M}+\mathcal{M}^\dagger U^\dagger e^{-i\theta/N_{\rm f}}            \right\},
\end{equation}
where $U\in SU(N_{\rm f})$ contains the pseudoscalar degrees of freedom and $\mathcal{M}$ is the mass matrix. For simplicity we consider a mass matrix proportional to the identity, $\mathcal{M}=m\mathbb{I}$.\\
$F$ and $\Sigma$ are the pseudoscalar decay constant and the quark condensate in the chiral limit,
and $\theta$ represents the vacuum angle.

In a finite volume $V=TL^3$ with $L\gg 1/\Lambda_{\rm QCD}$, one can distinguish different chiral regimes. 
If $M_P$ is the pseudoscalar meson mass, approaching the chiral limit by keeping $M_P L\gg 1$ defines the so-called $p$-regime. In this case the chiral effective theory looks essentially as in the infinite volume case: finite-volume effects are exponentially suppressed by factors $\exp{(-M_P L)}$, while the mass effects are the dominant ones. The power counting in terms of the momentum $p$ and quark mass $m$ is given by
\begin{equation}
m\sim p^2,\;\;\;\;1/L,1/T\sim p.
\end{equation}
Alternatively, one can approach the chiral limit while keeping $\mu=m\Sigma V\lesssim 1$; in this case the Compton wavelength associated with the pseudo-Goldstone bosons is much larger than the linear extent $L$ of the box, and volume effects are enhanced. This defines the $\epsilon$-regime, where the power-counting is reorganised such that \cite{Gasser:1986vb,Gasser:1987ah}
\begin{equation}
m\sim \epsilon^4,\;\;\;\;1/L,1/T\sim \epsilon.
\end{equation}
One of the most important effects of the reorganisation of the power counting is that, at a given order in the effective theory, fewer Low Energy Couplings (LECs) appear with respect to the $p$-expansion. The fact that the corresponding higher-order counterterms are
kinematically suppressed may be convenient for the extraction of LECs by matching the effective theory to lattice QCD.

In this work we consider the left-handed current, which at leading order in the effective theory formalism corresponds to
\begin{equation}
\mathcal{J}_\mu^a=\frac{F^2}{2}{\rm Tr}\left(T^a U \partial_{\mu}U^\dagger   \right),
\end{equation}
where $T^a$ are the traceless generators of SU$(N_{\rm f})$.
In particular we are interested in the two-point correlation function
\begin{equation}\label{2p_f}
\mathcal{C}^{ab}(t)=\int d^3 x \langle \mathcal{J}_0^a(x) \mathcal{J}_0^b(0)\rangle ={\rm Tr}[T^aT^b]\mathcal{C}(t).
\end{equation}
The chiral effective theory formalism can be extended to the quenched case; in particular, two equivalent methods have been developed to cancel the fermion determinant, namely the supersymmetric formulation and the replica method \cite{Bernard:1992mk,Sharpe:1992ft,Damgaard:2000gh}. An important feature of the quenched setup is that the flavour singlet does not decouple in this case; moreover, its mass parameter $m_0^2/(2N_c)$ is related to the topological susceptibility.\\
In the following we summarise the known results for this current correlator from quenched chiral perturbation theory at NLO\footnote{We neglect terms proportional to $\frac{\alpha}{N_c}$, which is the parameter associated to the kinetic term of the singlet field, since it is suppressed for this observable in the simultaneous expansion in momenta and $1/N_c$ \protect\cite{Kaiser:2000gs}.}, in the $\epsilon$- and $p$-regimes, with degenerate quark masses.
Current correlators have been recently computed in the effective theory also for non-degenerate quark masses, in the full and partially quenched scenarios, in the case where all quarks are in the $p$- or in the $\epsilon$-regime and in the mixed case, where $m_v\Sigma V\lesssim 1$ for the valence quarks and $m_s\Sigma V\gg 1$ for the sea quarks \cite{Bernardoni:2007hi,Damgaard:2007ep}. 
\subsection{$\epsilon$-regime}
Topology plays a relevant r\^ole in the $\epsilon$-regime \cite{Leutwyler:1992yt}, such that observables must be considered in sectors of fixed topological charge. 
In the quenched case, the current correlator in Eq.~(\ref{2p_f}) at NLO and fixed topology $\nu$ is given by \cite{Damgaard:2002qe, Hernandez:2002ds}
\begin{equation}\label{eq:ctnlo}
\mathcal{C}_{\nu}(t)=\frac{F^2}{2T}\left\{1+\frac{2\mu T^2}{F^2V} \sigma_\nu(\mu)h_1\left(\frac{t}{T}\right)  \right\},
\end{equation}
with 
\begin{equation}
h_1(\tau) =\frac{1}{2}\left[\left(|\tau|-\frac{1}{2}\right)^2-\frac{1}{12}          \right],
\end{equation}
and 
\begin{equation}\label{eq:cond}
\sigma_\nu(\mu)=\mu\left[I_\nu(\mu)K_\nu(\mu)+I_{\nu+1}(\mu)K_{\nu-1}(\mu)    \right]+\frac{\nu}{\mu},
\end{equation}
where $I_\nu$ and $K_\nu$ are modified Bessel functions.
The most notable fact is that in the NLO expression only the leading-order LECs $\Sigma$ and $F$ enter, as already anticipated. \\
In our analysis we will compare both the time and topology dependence of the QCD correlators computed on the lattice with the expectations of the effective theory.
A convenient way to study the topology dependence is to fix $t=T/2$; in the chiral limit one has
\begin{equation}
\mu\sigma_\nu(\mu)|_{\mu=0}=|\nu|,
\end{equation}
hence one expects $\mathcal{C}_\nu(T/2)$ to depend linearly on the topological charge $\nu$. Moreover, one obtains the parameter-free prediction
\begin{equation}\label{sumrule}
24L^3\left[\mathcal{C}_{\nu_1}(T/2)-\mathcal{C}_{\nu_2}(T/2)         \right]|_{\mu=0}=|\nu_2|-|\nu_1|.
\end{equation}
From this expression and from Eq.~(\ref{eq:ctnlo}) it becomes clear that the sensitivity to topology is quite limited: in order for $\mathcal{C}_{{\nu}}(T/2)$ to be
significantly different from $\mathcal{C}_{\nu+\Delta\nu}(T/2)$ one needs the following condition on the relative error:
\begin{equation}
\frac{\Delta\mathcal{C}_{\nu}(T/2)}{\mathcal{C}_{\nu}(T/2)}\ll\frac{\Delta\nu}{12(FL)^2}\frac{T}{L}.
\end{equation}
Using the quenched value $F\simeq 100$ MeV from \cite{Giusti:2004yp}, this implies that statistical errors much smaller than $\sim[(14\Delta\nu)T/L]\%$ for $L= 1.5$ fm and $\sim[(8\Delta\nu)T/L]\%$ for $L= 2.0$ fm must be reached.
Notice that NLO effects are larger for asymmetric boxes, since they are proportional to $(T/L)^3$;
however, if $T\gg L$ one enters in a different kinematic range, called $\delta$-regime, which will not be discussed in this work.\\
For $\mu\ll 1$, the leading $\mu$-dependence in the NLO correction is given by
\begin{equation}\label{sigma_mass}
\mu\sigma_\nu(\mu)=\left\{
\begin{array}{ll}
|\nu|+\frac{\mu^2}{2|\nu|} +...& (\nu\neq 0)\\
\left[\frac{1}{2}-\gamma-\log\left(\frac{\mu}{2}  \right)   \right]\mu^2+...& (\nu=0),
\end{array}
\right.
\end{equation}
where $\gamma$ is the Euler-Mascheroni constant.
For $\nu\neq 0$ we then expect a weak sensitivity to quark mass for the current correlator.

Matching the left correlator computed in lattice QCD with the chiral effective theory allows to extract the low-energy constant $F$ with control over NLO effects. For this particular correlator the chiral condensate $\Sigma$ appears only at NLO; in particular
the expression in Eq.~(\ref{eq:cond}) represents the quenched chiral condensate at finite $\mu$ at leading order in the $\epsilon$-expansion \cite{Osborn:1998qb,Damgaard:1998xy}.
At NLO, the condensate retains the same functional form of Eq.~(\ref{eq:cond}) \cite{Osborn:1998qb,Damgaard:2001js}, with $\mu$ replaced by $\mu_{\rm eff}=(m\Sigma_{\rm eff}V)$ and 
\begin{equation}\label{sigmaeff}
\Sigma_{\rm eff}(V)=\Sigma\left[1+w_0\bar{H}(0)\right],
\end{equation}
where 
\begin{equation}
\bar{H}(x)=\frac{1}{V}\sum_{p\neq 0}\frac{1}{(p^2)^2}\;e^{ipx}.
\end{equation}
Moreover, we have defined
\begin{equation}
w_0=\frac{m_0^2}{2N_c F^2},
\end{equation}
where $m_0^2/(2N_c)$ is the flavor singlet mass parameter;
as already anticipated, it is related to the topological susceptibility by the equation\footnote{We adopt the normalisation conventions of \protect\cite{Damgaard:2002qe,Hernandez:2002ds}.}
\begin{equation}\label{m0}
\frac{\langle \nu^2\rangle}{V}=\frac{m_0^2F^2}{4N_c}.
\end{equation}
In dimensional regularisation one obtains
\begin{equation}
\bar{H}(0)=\beta_2+\frac{1}{(4\pi)^2}\left[1+2c_1+\ln\left(\frac{\hat{L}^2}{L_0^2}  \right)   \right]
\end{equation}
where $\hat{L}=V^{1/4}$, $1/L_0$ is the ultraviolet subtraction point, and 
\begin{equation}
c_1=\frac{1}{4-d}+\frac{1}{2}\left(-\gamma+\ln(4\pi)   \right).
\end{equation}
$\beta_2$ is a shape coefficient \cite{Hasenfratz:1989pk}, which in the symmetric case $T=L$ takes the value
\begin{equation}
\beta_2=-0.020305.
\end{equation}
The infrared ``sickness'' of quenched QCD is reflected here in the fact that $\Sigma_{\rm eff}(V)$ diverges in the limit $\hat{L}\rightarrow \infty$. In the following we define operatively $\Sigma_{\rm eff}$ at a fixed $V$ as
\begin{equation}\label{sigmaeff1}
\left[\frac{2|\nu|\Sigma}{mV}\left(\sigma_\nu(\mu)-\frac{|\nu|}{\mu}\right)\right]_{m=0}\equiv\Sigma^2_{\rm eff}(L),\;\;\;\;\nu>0.
\end{equation}
\subsection{$p$-regime}
In the $p$-regime, the NLO finite-volume prediction for the current correlator in the quenched case is given by \cite{Colangelo:1997ch}
\begin{equation}\label{corr_que}
\mathcal C(t)=\frac{1}{2}M_P^V(F_P^V)^2 \frac{\cosh\left[(T/2-t)M_P^V\right]}{2\sinh\left[TM_P^V/2  \right]}.
\end{equation}
The pseudoscalar decay constant in this case is volume-independent:
\begin{equation}\label{eq:fp}
F_P^V=F_P=F\left[1+\frac{M^2}{2(4\pi F)^2}\alpha_5             \right],
\end{equation}
where $\alpha_i$ are the LECs associated with NLO operators in the (quenched) chiral Lagrangian in the convention of \cite{Heitger:2000ay}, and 
\begin{equation}\label{eq:msq}
  M^2=\frac{2m\Sigma}{F^2}.
\end{equation}
For the finite-volume pseudoscalar meson mass one obtains
\begin{eqnarray}
\left(M_P^{V}\right)^2 &= & M_P^2\left[1+w_0g_2(M_P,V)  \right],\label{mp_fv1}\\
M_P^2 & = & M^2\left[1+w_0H(M^2) -\frac{M^2}{(4\pi F)^2}(\alpha_5-2\alpha_8)           \right],\label{mp_fv2}
\end{eqnarray}
where $H(M^2)$ is given, in dimensional regularisation, by
\begin{equation}\label{h_dim}
H(M^2)=\int \frac{d^dp}{(2\pi)^d}\frac{1}{(p^2+M^2)^2}=\frac{1}{(4\pi)^2}\left[2c_1-\ln\left(\frac{M^2}{\mu^2} \right)   \right].
\end{equation}
The volume-dependent function $g_r$ reads \cite{Hasenfratz:1989pk}
\begin{equation}
g_r(M_P,V)=\frac{1}{(4\pi)^2}\int_0^\infty \frac{d\lambda}{ \lambda^{3-r}}e^{-\lambda M_P^2}\sum_{n\in\mathbb{Z}^4}\left(1-\delta_{n,0}^{(4)} \right)\times
\end{equation}
$$
\exp\left[-\frac{1}{4\lambda}\left( T^2n_0^2+L^2\sum_{i=1}^3n_i^2      \right)  \right].
$$
In our analysis we will investigate finite-volume effects by comparing lattice results obtained at different volumes $V_1$ and $V_2$. A convenient quantity to consider for this purpose is the ratio $M_P^{V_1}/M_P^{V_2}$, for which one obtains the NLO expression 
\begin{equation}\label{eq:mp_ratio}
\left(\frac{M_P^{V_1}}{M_P^{V_2}}\right)^2=1+w_0\Big[g_2(M_P,V_1)-g_2(M_P,V_2) \Big].
\end{equation}
If $w_0$ is given as input, this is a parameter-free prediction from the chiral effective theory.\\
By reabsorbing the divergences in Eq.~(\ref{h_dim}) in the low-energy constant $\Sigma$ one obtains
\begin{equation}
\frac{M_P^2}{2m}=\frac{\Sigma(\mu)}{F^2}\left[1-\frac{w_0}{(4\pi)^2}\log\left(\frac{M^2}{\mu^2} \right)-   \frac{M^2}{(4\pi F)^2}(\alpha_5-2\alpha_8)         \right].
\end{equation}
$\Sigma(\mu)$ is related at NLO to the $\Sigma_{\rm eff}$ at a given scale $(L_{\rm eff})$ defined in the $\epsilon$-regime in Eq.~(\ref{sigmaeff1}) by 
\begin{equation}\label{sigma_rel}
\Sigma_{\rm eff}(L_{\rm eff})=\Sigma(\mu)\left[1+w_0\left(\beta_2+\frac{1}{(4\pi)^2}\left(1+\log(L_{\rm eff}^2\mu^2)\right)\right)   \right].
\end{equation}

\section{Numerical simulations}\label{sec3}
For our numerical study we have adopted the Neuberger-Dirac operator $D$ \cite{Neuberger:1997fp,Neuberger:1998wv} and the Wilson gauge action, on a box with volume $V=L^3T$ and lattice spacing $a$.
We have computed the two-point function 
\begin{equation}
C^{ab}(t)=\sum_{\vec{x}}\langle J_0^a(x)J_0^b(0)\rangle={\rm Tr}[T^a T^b]C(t),
\end{equation}
with the left-handed current
\begin{equation}
J_0^a=\overline\psi T^a\gamma_0P_-\widetilde{\psi},
\end{equation}
where $P_{\pm}=(1\pm\gamma_5)/2$ and
\begin{equation}
\widetilde{\psi}=\left(1-\frac{1}{2}\overline{a} D\right)\psi,\;\;\;\overline{a}=\frac{a}{1+s}.
\end{equation}
The parameter $|s|<1$ has been chosen equal to 0.4. For complete definitions and conventions related to the Neuberger-Dirac operator the reader can refer to \cite{Giusti:2004yp}. An advantage of using left-handed currents is that zero-modes of the Dirac operator do not contribute to the corresponding correlator: at finite volume, no divergences are present in the $m\rightarrow 0$ limit.\\
We have considered two sets of lattices, one dedicated to simulations in the $p$-regime and one for the $\epsilon$-regime. The parameters of the two sets are reported in Tables \ref{tab_p} and \ref{tab_eps}.
For the $p$-regime we have chosen two different lattice spacings and two spatial extents $L\simeq 1.5$ fm (p1 and p3) and $L\simeq 2$ fm (p2). The temporal extent is chosen to be $T=2L$ (p1,p2) or $T=3L/2$ (p3). For the lattice p3 we did not perform new simulations and we used instead the data already presented in \cite{Giusti:2004yp}. \\
For the $\epsilon$-regime, we have chosen three symmetric lattices (e1, e2, e3) with the same parameters used in a previous work \cite{Giusti:2007cn} for the computation of the quark condensate.
The quark masses here have been chosen such that $(mV)/(Z_Sr_0^3)$ is constant for the lattices (e1, e2, e3), where $Z_S$ is the renormalisation constant of the RGI scalar density \cite{Hernandez:2001yn,Wennekers:2005wa} and $r_0\simeq 0.5$ fm \cite{Sommer:1993ce}. Moreover, we have considered two additional asymmetric lattices, e4 and e5, with $T=2L$. For the lattice e4 we used the measurements collected in a previous project \cite{Giusti:2006mh}.  \\
Following \cite{DeGrand:2004qw,Giusti:2004yp}, we applied the low-mode averaging technique in order to reduce large fluctuations induced by low-modes wave functions. In Tables \ref{tab_p} and \ref{tab_eps}, $N_{\rm low}$ indicates the number of low-modes which have been extracted. The computation of the topological index, the low-lying eigenvalues and the inversion of the Neuberger Dirac operator have been performed using the techniques in \cite{Giusti:2002sm}. 
For all the lattices, the low eigenvalues have been computed with a $5\%$ precision. The values of the scale $r_0$ in lattice units \cite{Necco:2001xg}, $Z_S$, and the renormalization factor for the local left-handed current \cite{Wennekers:2005wa} for our couplings $\beta$ are listed in Table \ref{tab_r0Z}.
\begin{table}[h]
\begin{center}
\begin{tabular}{|c| c c |c| c| c|}
\hline
lat  &  $\beta$   &   $V$   &  $N_{\rm cfg}$   & $am$ & $N_{\rm low}$\\
\hline
p1   &  5.8458    &  $12^3\times 24$ &  475   & 0.02, 0.03, 0.04, 0.06 & 7 \\
\hline
p2   &  5.8485    &  $16^3\times 32$  & 197     &   0.02, 0.03, 0.04, 0.06 & 20 \\    
\hline
p3   & 6.0        &  $16^3 \times 24$ & 113  &  0.025, 0.04,0.06,0.08,0.1 & 8 \\
\hline
\end{tabular}
\caption{Simulation parameters for the $p$-regime}\label{tab_p}
\end{center}
\end{table}
\begin{table}[h]
\begin{center}
\begin{tabular}{|c| c c |c| c| c|}
\hline
lat  &  $\beta$   &   $V$   &  $N^{\nu}_{\rm cfg}$   & $am$ & $N_{\rm low}$ \\
\hline
e1   &  5.8458    &  $12^4$ &  177, 313,221,   & 0.001, 0.003, 0.008, & 20 \\
     &            &         &  126,62,35       & 0.012,0.016        &  \\
\hline
e2   &  5.8458    &  $16^4$ & 49,69,82,72     &                   0.000316, 0.000949,0.00253, & 20\\
     &            &         & 50,54,38,32      &                  0.00380,0.00506 & \\
\hline
e3   & 6.0        &  $16^4$ & 131,231,178,  &                      0.000612, 0.00184,0.00490, & 20 \\
     &            &         & 96,44,16      &                  0.00735, 0.00980 & \\
\hline
e4   & 5.8485     &  $16^3\times 32$      & 151, 130, 125, 101,87 &    0.002,0.003 & 20  \\
     &            &                       & 86,66,52, 29          &              &   \\
\hline
e5   & 5.8458     &  $12^3\times 24$      & 22, 55, 50, 56,25 &    0.003 & 7 \\
     &            &                       & 21,22          &             &    \\
\hline
\end{tabular}
\caption{Simulation parameters for the $\epsilon$ regime. $N^{\nu}_{\rm cfg}$ indicates the number of configurations for topologies $|\nu|=0,1,2...$, except for lattice e4, where the lowest index is $|\nu|=2$.}\label{tab_eps}
\end{center}
\end{table}
\begin{table}[h]
\begin{center}
\begin{tabular}{|c |c |c|c| }
\hline
$\beta$  &  $r_0/a$  &  $Z_J$ &  $Z_S$ \\
\hline
5.8458   &  4.026(23)         &  1.710(5) & 1.28(6)    \\
5.8485   &  4.048(23)         &  1.706(5) & 1.28(6)\\
6.0      &  5.37(3)         &  1.553(2) & 1.05(5)      \\
\hline
\end{tabular}
\caption{Values of the Sommer scale $r_0/a$ \protect\cite{Necco:2001xg} and the renormalisation constants of the left-handed current and the scalar density \protect\cite{Wennekers:2005wa} for the values of $\beta$ used in our study.} \label{tab_r0Z}
\end{center}
\end{table}

\section{Matching lattice QCD with the chiral effective theory: 
$\epsilon$-regime}\label{sec4}
\subsection{LO matching}
\begin{figure}
\begin{center}
\includegraphics[width=9cm,angle=-90]{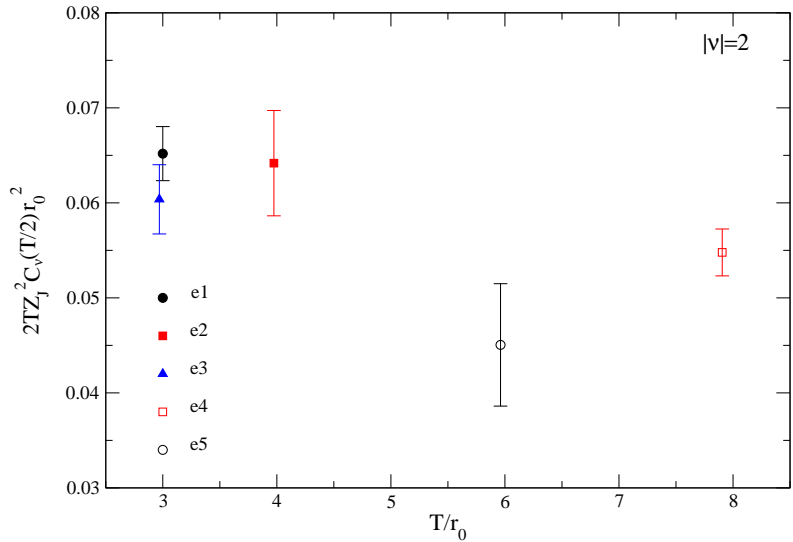}
\caption{The quantity $2T Z_J^2  C_\nu(T/2)r_0^2$ plotted as a function of $T/r_0$ for lattices e1 (black filled circles), e2 (red filled squares) and e3 (blue filled triangles).
The plot reports the data at fixed topological charge $|\nu|=2$ and fixed $\mu\simeq 0.6$, corresponding to the intermediate quark mass. 
The red empty squares correspond to the lattice e4 ($\mu\simeq 0.5$, corresponding to the lightest mass), while the black empty circles correspond to the lattice e5 ($\mu\simeq 1$).} \label{fig:eps_lo}
\end{center}
\end{figure}

At LO in the $\epsilon$-expansion, the left correlation function is expected to be independent on time, mass and topology; from Eq.~(\ref{eq:ctnlo}) one reads
\begin{equation}
\mathcal{C}_\nu(t)= \frac{F^2}{2T}.
\end{equation}
In Fig.~\ref{fig:eps_lo} we plot the renormalised dimensionless quantity $2T Z_J^2 C_\nu(T/2) r_0^2$ as a function of $T/r_0$, for $|\nu|=2$ .
We do not observe a significant mass dependence of this quantity, hence we report the results only at a single value of $\mu$ (see figure caption). \\
The results obtained for lattices e1 (black filled circles) and e3 (blue filled triangles) corresponding to $T\simeq 3r_0$ are in good agreement between each other; this indicates that within our precision we are not sensitive to lattice artifacts. 
Their consistency with the data of lattice e2 (red filled squares), for which $T\simeq 4r_0 $, is a verification of the $1/2T$ scaling predicted at LO.
In this plot we also show the data obtained with the asymmetric lattices e4 (for $am=0.002$, corresponding to $\mu\simeq 1$) and e5 ($\mu\simeq 0.5$). In this case the data tend to depart from the results of the symmetric lattices, although not significantly within our statistical errors. From the chiral effective theory we expect NLO effects to be larger for asymmetric volumes, as already discussed in section \ref{sec2}.
\subsection{NLO matching: topology dependence}
\begin{figure}
\begin{center}
\includegraphics[width=10cm,angle=-90]{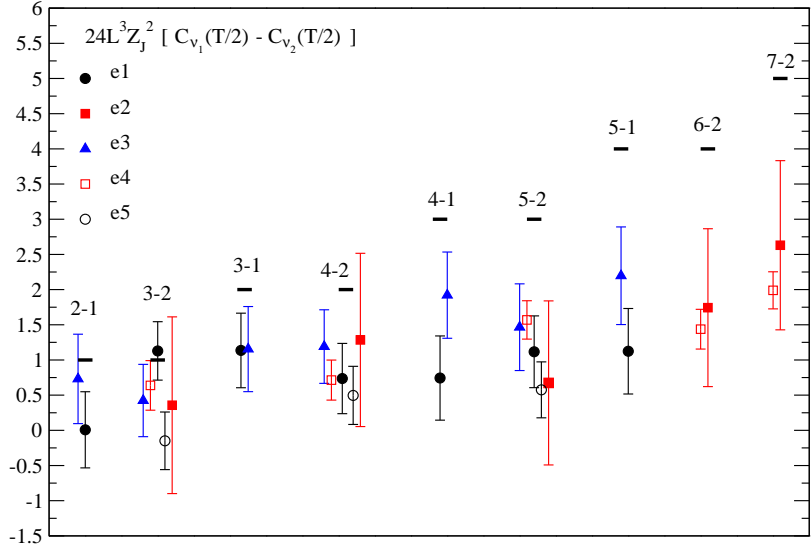}
\caption{The differences $24L^3Z_J^2\left[C_{\nu_1}(T/2)-C_{\nu_2}(T/2)  \right]$ computed in the $\epsilon$-regime.
The lines represent the parameter-free prediction from the chiral effective theory (Eq.~(\protect\ref{sumrule})), and the numbers denote the different combinations of $(|\nu_2|-|\nu_1|)$. The data from lattices e1, e2 and e3 refer to $\mu\simeq 0.6$, corresponding to the intermediate quark mass. For the lattice e4, the lightest quark mass $\mu\simeq 0.5$ is considered.
}\label{fig:eps_nlo}
\end{center}
\end{figure}

At NLO, the $\epsilon$-expansion gives predictions on time, mass, and topology dependence of $C_{\nu}(t)$; in particular,
one expects $C_{\nu}(T/2)$ at the chiral limit to depend linearly on the topological charge.\\
In Fig.~\ref{fig:eps_nlo} we show the differences $24L^2Z_J^2\left[C_{\nu_1}(T/2)-C_{\nu_2}(T/2)  \right]$ for several choices of $\nu_1,\nu_2$  and compare the results with the parameter-free predictions of the chiral effective theory, Eq.~(\ref{sumrule}).
For the lattices e2 and e5 we excluded the sector $|\nu|=1$ because of large statistical uncertainties. 
As already noticed in the previous section, we don't observe significant quark mass dependence.
The statistical errors associated with the sector $\nu=0$ are very large.
 Hence, also in the case where we would expect a stronger quark mass
 dependence as predicted by Eq.~(\protect\ref{sigma_mass}), we are not sensitive to mass
 effects.
Anyway we exclude the $\nu=0$ sector from this analysis.\\
Numerical data tend to depart systematically from the theoretical expectations when the topological
charge is increased.
Although statistical uncertainties  associated with the results are too large for precise quantitative statements, this may indicate that higher order corrections are significant.
\subsection{NLO matching: time dependence and determination of $F$.}
\begin{figure}
\begin{center}
\includegraphics[width=8cm, angle=-90]{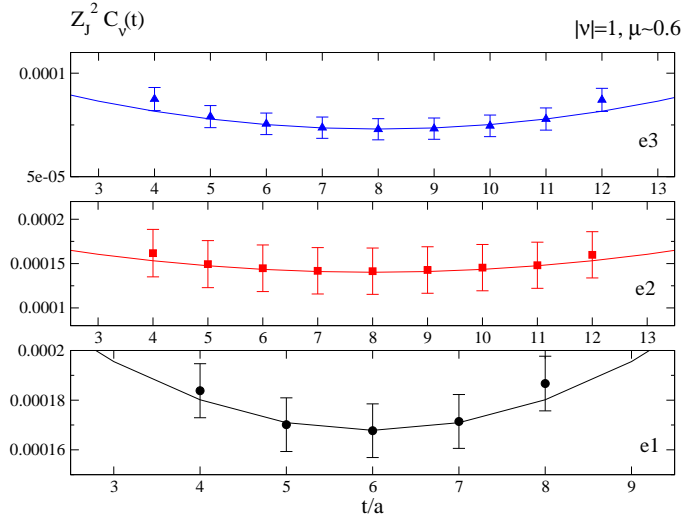}  
\caption{Time dependence of the correlator $Z_J^2C_\nu(t)$ for $|\nu|=1$ at the intermediate quark mass, for the lattices e1, e2, e3. The curves represent the fit of Eq.~(\protect\ref{eq:fit1}), with ranges $t/a=5-7$ (e1), $t/a=5-11$ (e2) and $t/a=6-10$ (e3).}\label{fig:ct}
\end{center}
\end{figure}

At fixed values of the topological charge and quark mass, the $\epsilon$-expansion predicts a parabolic time dependence of current correlators.
Since we have already observed that our sensitivity is not high enough to test NLO dependence, we do not enforce the topology and mass dependence of the NLO corrections and leave it as a free parameter. In particular we perform the 2-parameter fit
\begin{equation}\label{eq:fit1}
Z_J^2 C_\nu(t)=\frac{B_1^2}{2T}+\frac{1}{L^3}B_2h_1\left(\frac{t}{T}\right).
\end{equation}
The results for $B_1$, $B_2$ at the different masses and topological charges are collected in the Appendix \ref{app_num} in Tables \ref{e1_res}, \ref{e2_res}, \ref{e3_res}. Three examples of the fit are given in Fig.~\ref{fig:ct}, for the intermediate mass and $|\nu|=1$.
The fit ranges are $t/a=5-7$, $t/a=5-11$ and $t/a=6-10$ for the lattices e1, e2 and e3 respectively. Asymmetric lattices are not considered in this analysis.
We observe that the results for $B_1$ are stable within errors with respect to the quark mass and topology. Moreover, each topological sector yields an independent determination and it is possible to reduce the associated error by performing averages between different sectors. 
By varying the fit range, no significant deviation is observed for $B_1$.
We determine $F$ in lattice units by averaging the results obtained for $B_1$ at the five quark masses, and then performing an additional average over different topological sectors. In particular, for the lattices e1, e3 we consider $|\nu|=1-3$, while for the lattice e2 we averaged in the interval $|\nu|=2-4$. This choice is due to large fluctuations which affect the sector $|\nu|=1$; the same happens for the $\nu=0$ sector, for all lattices. For the dimensionless quantity $Fr_0$ we obtain
\begin{eqnarray}
Fr_0 & = & 0.284(4),\;\;\;{\rm (e1)} \nonumber\\
Fr_0 & = & 0.278(6), \;\;\;{\rm (e2)}\label{Fres_eps1}\\
Fr_0 & = & 0.280(5).  \;\;\;{\rm (e3)}\nonumber
\end{eqnarray}
The agreement between (e1) and (e2) indicates that finite-volume effects are below our statistical precision; their agreement with (e3) is a signal that also lattice artifacts are smaller than our errors.

\begin{figure}
\begin{center}
\includegraphics[width=9cm,angle=-90]{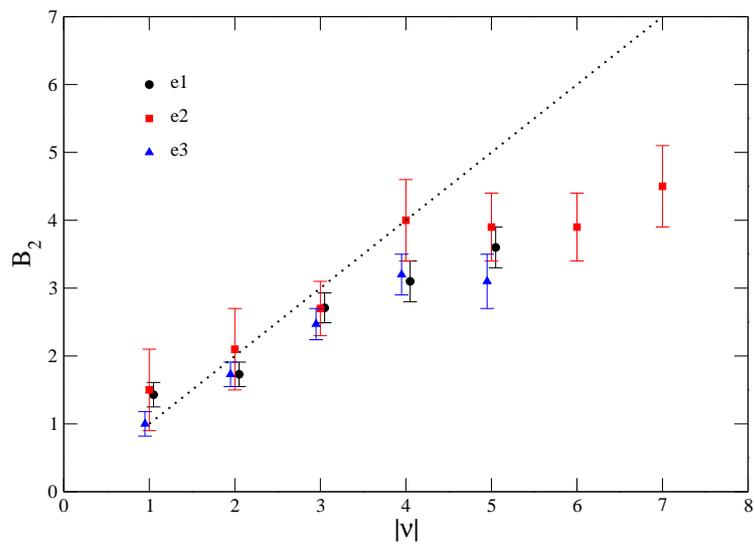}
\caption{The coefficient $B_2$ of Eq.~(\protect\ref{eq:fit1}) as a function of the topological charge $|\nu|$. The data refer to the smallest quark mass. The results for the lattices e1 and e3 are slightly shifted on the horizontal axis.
The dotted line represents the theoretical expectation from the chiral effective theory, $B_2=|\nu|$. }\label{fig:b2_nu}
\end{center}
\end{figure}

\begin{figure}
\begin{center}
\includegraphics[width=9cm,angle=-90]{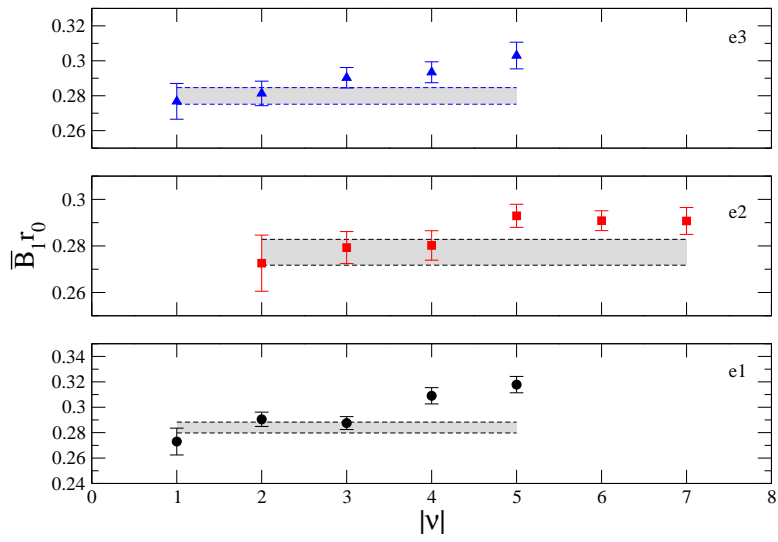}
\caption{The coefficient $\overline{B}_1$ of the constrained fit Eq.~(\protect\ref{eq:fit2}) as a function of the topological charge $|\nu|$. The data refer to the smallest quark mass. The grey bands represent the results of Eq.~(\protect\ref{Fres_eps1}). For the lattice e2, results for $|\nu|=1$ have been omitted due to large statistical errors. }\label{fig:b1_nu}
\end{center}
\end{figure}

In the chiral limit, the effective theory predicts the coefficient $B_2$ to be equal to the topological index $|\nu|$; in Fig.~\ref{fig:b2_nu} we report the results for $B_2$ at the smallest quark mass, as a function of $|\nu|$, together with the theoretical expectation. 
The results for $B_2$ are more sensitive to a change of the fit range with respect to $B_1$; in particular for the lattice e1, enlarging the fit range gives differences of the order of 2-3 standard deviations. This is not surprising, since the latter has a relatively coarse lattice spacing and a relatively small volume, and the  number of points available for the fit is hence limited. Here we consider only the results concerning the fit range $t/a=5-7$, keeping in mind that additional systematic error on $B_2$ for lattice e1 can be substantial.\\
Like in the previous analysis at fixed $t=T/2$, 
the data show a tendency to depart from the prediction with increasing $|\nu|$, indicating that higher order effects may be important. In particular, corrections are expected to be severe when $|\nu|\gg \sqrt{\langle \nu^2 \rangle}$.
However, also in this case the statistical errors are very large and do not allow for a precise statement.\\
In this fit, the stability of the coefficient $B_1$ with respect to $m$ and $\nu$ can be interpreted as a signal that systematic uncertainties coming from higher orders are under control. 
As additional test we can constrain the NLO term\footnote{The very weak dependence on the quark mass for $\nu\neq 0$ allows us to fix $B_2$ to the expected value in the chiral limit.} $B_2=|\nu|$ 
\begin{equation}\label{eq:fit2}
Z_J^2 C_\nu(t)=\frac{\overline{B}_1^2}{2T}+\frac{|\nu|}{L^3}h_1\left(\frac{t}{T}\right).
\end{equation}
The results obtained for $\overline{B}_1$ for different $|\nu|$ are shown in Fig.~\ref{fig:b1_nu} for the smallest quark mass; 
compatible results are obtained for the heavier quark masses. In this case the importance of higher order terms in the chiral theory manifests itself in the fact that $\overline{B}_1$ is topology dependent . However, for low $|\nu|$ we observe a constant behaviour within statistical errors: 
moreover, data at low $|\nu|$ are fully compatible with the results reported in Eq.~(\ref{Fres_eps1}) obtained from the unconstrained fit, which are represented by the grey band in the plot. 
This makes us confident that systematic errors on $F$ coming from higher order chiral corrections are smaller than the statistical uncertainties quoted in Eq.~(\ref{Fres_eps1}). Moreover, it also justifies the topology range that has been chosen at that stage to perform the average.



\section{Matching lattice QCD with the chiral effective theory: $p$-regime}\label{sec5}
The data obtained for the current correlator $C(t)$ in the $p$-regime can be compared with the expectation of the chiral effective theory at NLO, Eq.~(\ref{corr_que}); after symmetrising the correlator around $t=T/2$, we compute the effective pseudoscalar mass $aM_{P,{\rm eff}}^V(t)$ by solving the equation
\begin{equation}
\frac{C(t)}{C(t+1)}=\frac{\cosh\left[\left(T/2-t \right) M_{P,{\rm eff}}^V(t)    \right]}{\cosh\left[\left(T/2-t -1\right) M_{P,{\rm eff}}^V(t)    \right]}.
\end{equation}
From $aM_{P,{\rm eff}}^V(t)$ we then compute an effective pseudoscalar decay constant $aF_{P,{\rm eff}}^V(t)$
\begin{equation}
F_{P,{\rm eff}}^V(t)=2Z_J\sqrt{\frac{C(t)\sinh\left(M_{P,{\rm eff}}^V(t)T/2\right)    }{M_{P,{\rm eff}}^V(t)\cosh\left[\left(T/2-t \right) M_{P,{\rm eff}}^V(t)    \right]  }            }.
\end{equation}
The results for $aM_P^V$ and  $aF_P^V$ are obtained from a plateau (for $t\ge t_{\rm min}$)  and are given in Table \ref{tab_pres}.  
\begin{table}
\begin{center}
\begin{tabular}{|c| c |c |c|c|}
\hline
lat. &       $am$    & $aM_P^V$    &   $aF_P^V$ &  $t_{\rm min}/a$ \\
\hline
p1   &       0.02    & 0.209(4)            &  0.0702(8)   &  6        \\
     &       0.03    & 0.239(3)            &  0.0715(8)   &  6      \\
     &       0.04    & 0.267(3)            &  0.0727(7)   &  6      \\
     &       0.06    & 0.315(3)            &  0.0751(7)   &  6       \\
\hline
p2   &       0.02    & 0.197(3)            &  0.0695(6)    &  7      \\
     &       0.03    & 0.2308(24)            & 0.0707(7)   &  7       \\
     &       0.04    & 0.2604(23)            & 0.0719(7)   &  7       \\
     &       0.06    & 0.3114(23)            & 0.0742(8)   &  7        \\
\hline
p3   &       0.025    & 0.199(6)            & 0.0530(10)    &  6       \\
     &       0.04    &  0.242(5)           &  0.0551(10)    &  6     \\
     &       0.06    &  0.292(5)           &  0.0580(10)    &  6     \\
     &       0.08    &  0.335(4)           &  0.0609(10)    &  6      \\
     &       0.1    &   0.375(4)          &   0.0637(10)    &  6     \\
\hline
\end{tabular}
\caption{Results for meson masses and decay constants in lattice units computed in the $p$-regime. $t_{\rm min}/a$ indicates the first point in the plateau of 
 $aM_{P,{\rm eff}}^V(t)$ and $aF_{P,{\rm eff}}^V(t)$.}\label{tab_pres}
\end{center}
\end{table}
\subsection{Pseudoscalar decay constant}
\begin{figure}
\begin{center}
\includegraphics[width=9cm,angle=-90]{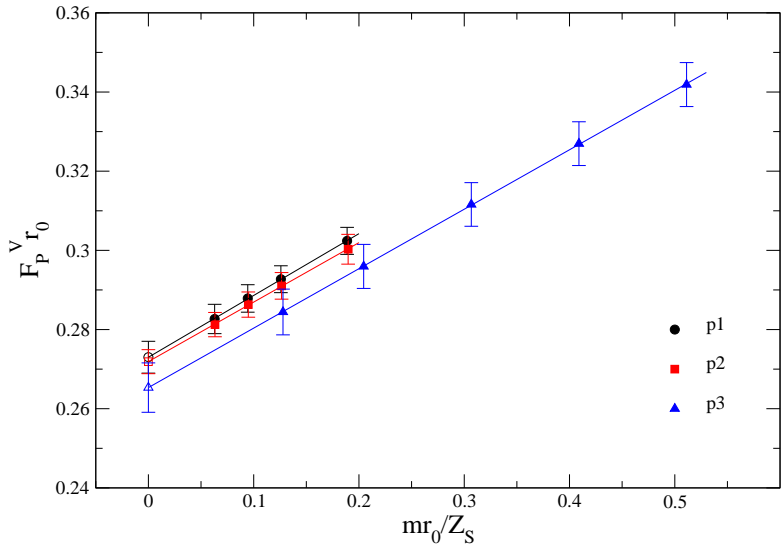}
\caption{ $F_P^Vr_0$ as a function of the quark mass $(mr_0/Z_S)$ for the lattices p1, p2, p3. }\label{fig:fpi_m}
\end{center}
\end{figure}

The quenched chiral effective theory predicts  $F_P^V$ to be volume-independent at NLO (see Eq.~(\ref{eq:fp})).
The volumes of the lattices p1 and p2 differ by a factor $\simeq 3$; 
for each quark mass, the results obtained for $aF_P^V$ for the two lattices are compatible within the statistical errors \footnote{The correction due to slightly different values of $\beta$ are smaller than the statistical uncertainty.}.\\
The quark mass-dependence of  $aF_P^V$ is well consistent with a linear behaviour.
A chiral extrapolation of the form
\begin{equation}\label{fp_linm}
aF_P=C_1+C_2 (am)
\end{equation}
yields for the three lattices the following results:
\begin{eqnarray}
Fr_0 & = & 0.273(4),\;\;{\rm (p1)}\nonumber   \\
Fr_0 & = & 0.272(3),\;\;{\rm (p2)}\label{Fr0_p1}   \\
Fr_0 & = & 0.265(6).\;\;{\rm (p3)}\nonumber 
\end{eqnarray}
In Fig.~\ref{fig:fpi_m} we show the quantity $F_P^Vr_0$ as a function of the renormalised quark mass $mr_0/ {Z}_S$ and the chiral extrapolation.
The chirally extrapolated values are in good agreement with each other, indicating that also lattice artifacts are below our statistical errors.
Alternatively, one can fit $F_P^V$ linearly in $(M_P^V)^2$, obtaining the LEC $\alpha_5$ directly from the slope:
\begin{eqnarray}
Fr_0 =0.268(4), & \alpha_5 = 1.83(13), & \;\;{\rm (p1)}\nonumber    \\
Fr_0 =0.269(3), & \alpha_5 = 1.69(14), & \;\;{\rm (p2)}\label{Falpha5_p1}  \\
Fr_0 =0.263(6), & \alpha_5 = 1.64(8).  & \;\;{\rm (p3)}\nonumber   
\end{eqnarray}
The extrapolated values for $Fr_0$ do not differ significantly from the previous ones. The values obtained for $\alpha_5$ are consistent with each other and with previous estimates \cite{Bardeen:2003qz, Giusti:2004yp}. The origin of the discrepancy with the results of \cite{Heitger:2000ay} have been discussed in \cite{Giusti:2004yp}.

\subsection{Pseudoscalar mass}
\begin{figure}
\begin{center}
\includegraphics[width=9cm,angle=-90]{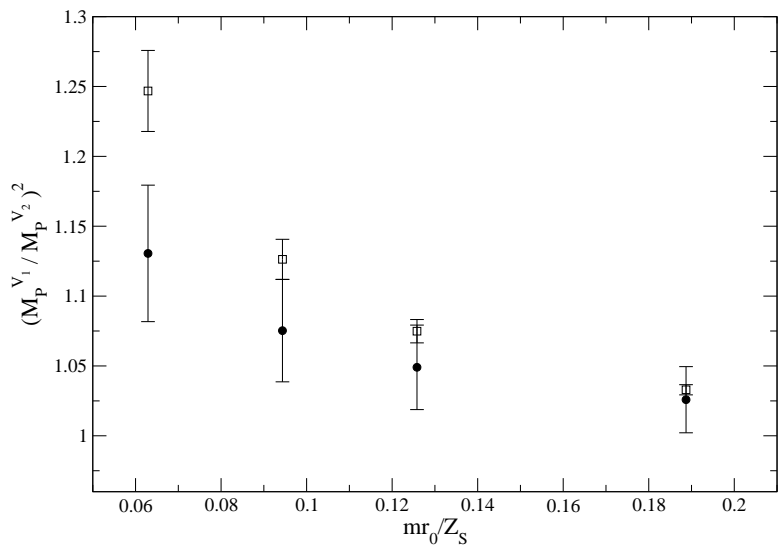}
\caption{Ratio of pseudoscalar masses, $(M_P^{V_1}/M_P^{V_2})^2$, where $V_1$ corresponds to the lattice p1 and $V_2$ to the lattice p2. The filled circles are the data; the empty squares represent the expectation from the chiral effective theory, Eq.~(\protect\ref{eq:mp_ratio}).}\label{fig:mp_ratio}
\end{center}
\end{figure}

\begin{figure}
\begin{center}
\includegraphics[width=8cm, angle=-90]{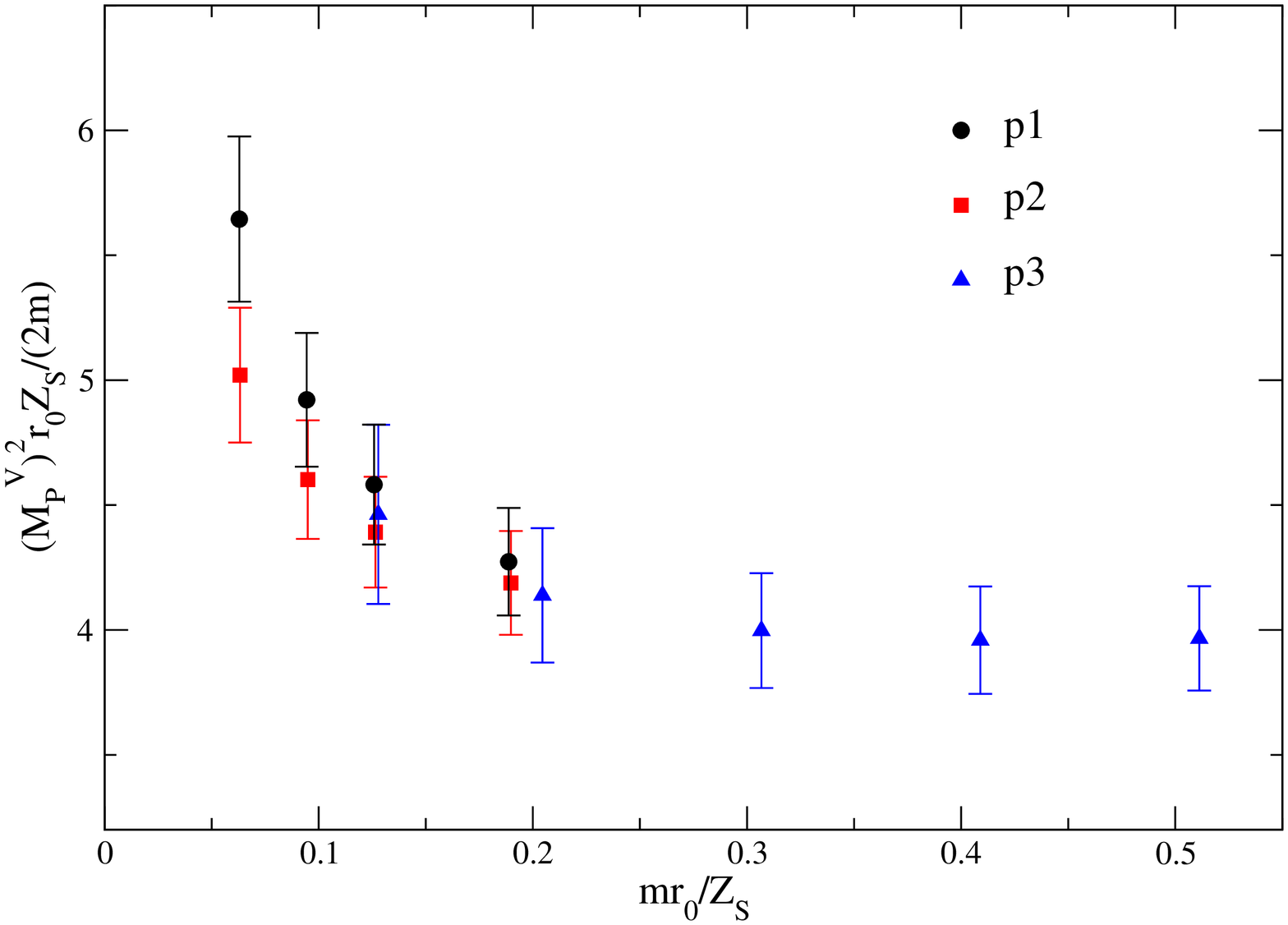}
\includegraphics[width=8cm, angle=-90]{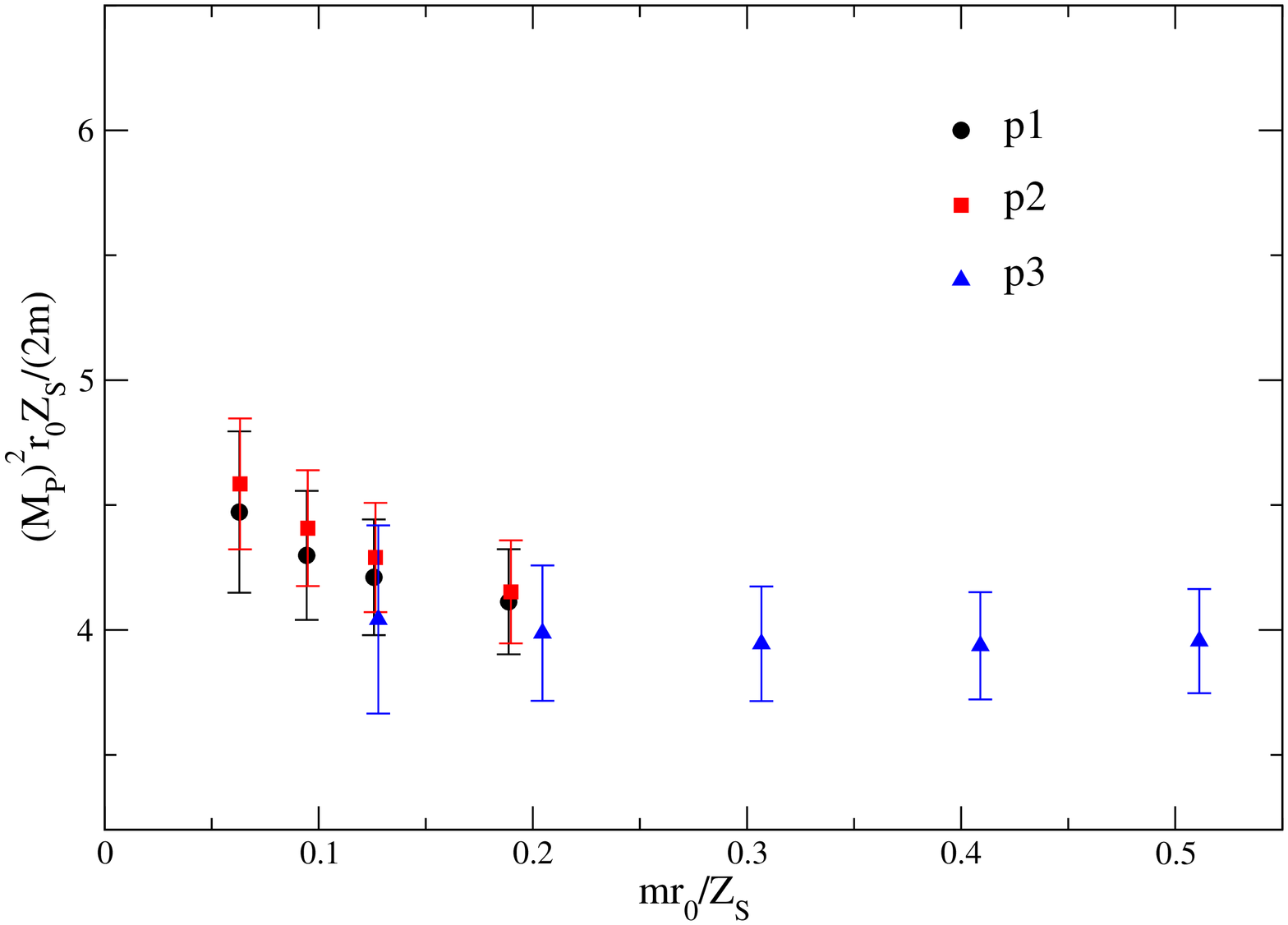}
\caption{Quark mass dependence of the squared pseudoscalar mass for lattices p1, p2, p3. The plot on the top shows the renormalised quantity  $\frac{(M_P^V)^2}{2m}r_0 Z_S$; on the bottom, finite volume corrections according to Eq.~(\protect\ref{mp_fv1}) are applied.} \label{fig:mp}
\end{center}
\end{figure}

We consider the ratio of pseudoscalar meson masses obtained with the lattices p1 and p2, which have different physical volume ($V_1$ and $V_2$ in the following). This ratio can be compared with Eq.~(\ref{eq:mp_ratio}),
using the topological susceptibility from \cite{DelDebbio:2004ns} and $Fr_0=0.265(6)$ from Eq.~(\ref{Fr0_p1}) for the lattice p3, that is
\begin{equation}\label{eq:w0}
w_0=\frac{m_0^2}{2N_cF^2}=23.9(2.5).
\end{equation}
 $M_P$ is substituted by $M_P^{V_2}$, corresponding to the largest volume; once $m_0$ and $F$ are given, Eq.~(\ref{eq:mp_ratio}) represents then a parameter-free prediction.
The results are reported in Fig.~\ref{fig:mp_ratio} (filled circles) for the different quark masses; the empty squares represent the parameter-free expectation from the chiral effective theory. The data are compatible with the NLO predictions; even though errors are too large to allow
for a precise comparison, NNLO corrections could be significant in this ratio.
On the basis of this analysis, we can correct our data for finite volume effects according to Eq.~(\ref{mp_fv1}) and then study the quark mass dependence of
$M_P^2/(2m)$. The results are presented in Fig.~\ref{fig:mp}, where we show the data before (top) and after (bottom) the finite volume corrections. Once the corrections are applied, $M_P^2/(2m)$ as a function of the quark mass is well compatible with a constant behaviour, without clear evidence for higher order effects.
One can use anyway the knowledge on NLO chiral effective theory and perform a two-parameter fit of the form
\begin{equation}
\frac{aM_P^2}{2m}=E_1\left[1-\frac{w_0}{(4\pi)^2}\log\left(\frac{M_P^2}{\mu^2}  \right)\right]+E_2M_P^2,
\end{equation}
with $w_0$ and $\mu$ given as external inputs. From $E_1$ and $E_2$ we obtain
\begin{equation}
\Sigma(\mu)=E_1F^2;\;\;\;\;\;\;2\alpha_8-\alpha_5=(4\pi F)^2\frac{E_2}{E_1}.
\end{equation}
Using the results for $F$ from the linear chiral fit in $m$ as discussed in the previous section, we obtain for the chiral condensate
\begin{eqnarray}\label{sigma_preg}
1/\mu= 1.5\;{\rm fm}& : & \Sigma(\mu)r_0^3 = 0.42(4),\;\; {\rm (p1)} \nonumber\\
                  &   &  \Sigma(\mu)r_0^3 = 0.43(3),\;\; {\rm (p2)}\nonumber\\
                  &   &  \Sigma(\mu)r_0^3 = 0.41(5),\;\; {\rm (p3)}\nonumber\\
1/\mu= 2.0\;{\rm fm}& : & \Sigma(\mu)r_0^3 = 0.47(4),\;\; {\rm (p1)} \\
                  &   &  \Sigma(\mu)r_0^3 = 0.48(4),\;\; {\rm (p2)}\nonumber\\
                  &   &  \Sigma(\mu)r_0^3 = 0.47(6).\;\; {\rm (p3)}\nonumber
\end{eqnarray}
The combination $(2\alpha_8-\alpha_5)$ can be extracted with very limited accuracy; from this fit we obtain errors of order 30-40 $\%$. A more realistic estimation of the errors
can be made by considering a double-ratio fit
\begin{equation} 
\frac{\displaystyle{\left(\frac{M_P^2}{2m}\right)}}{\displaystyle{\left(\frac{M_{P{\rm ref}}^2}{2m_{\rm ref}}\right)}}=1-\frac{w_0}{(4\pi)^2}\log\left(\frac{M_P^2}{M_{P{\rm ref}}^2}\right)+
\bar{E}_2(M_P^2-M_{P{\rm ref}}^2),
\end{equation}
where one fixes a reference mass $m_{\rm ref}$ and one identifies $\bar{E}_2=2\alpha_8-\alpha_5$. From this one-parameter fit we obtain uncertainties of the order 50-80 $\%$  on  $(2\alpha_8-\alpha_5)$. It is then clear that our precision is too poor to quote a result for this combination.

\section{Comparing results from $\epsilon$- and $p$- regimes}\label{sec6}
\begin{figure}
\begin{center}
\includegraphics[width=8cm,angle=-90]{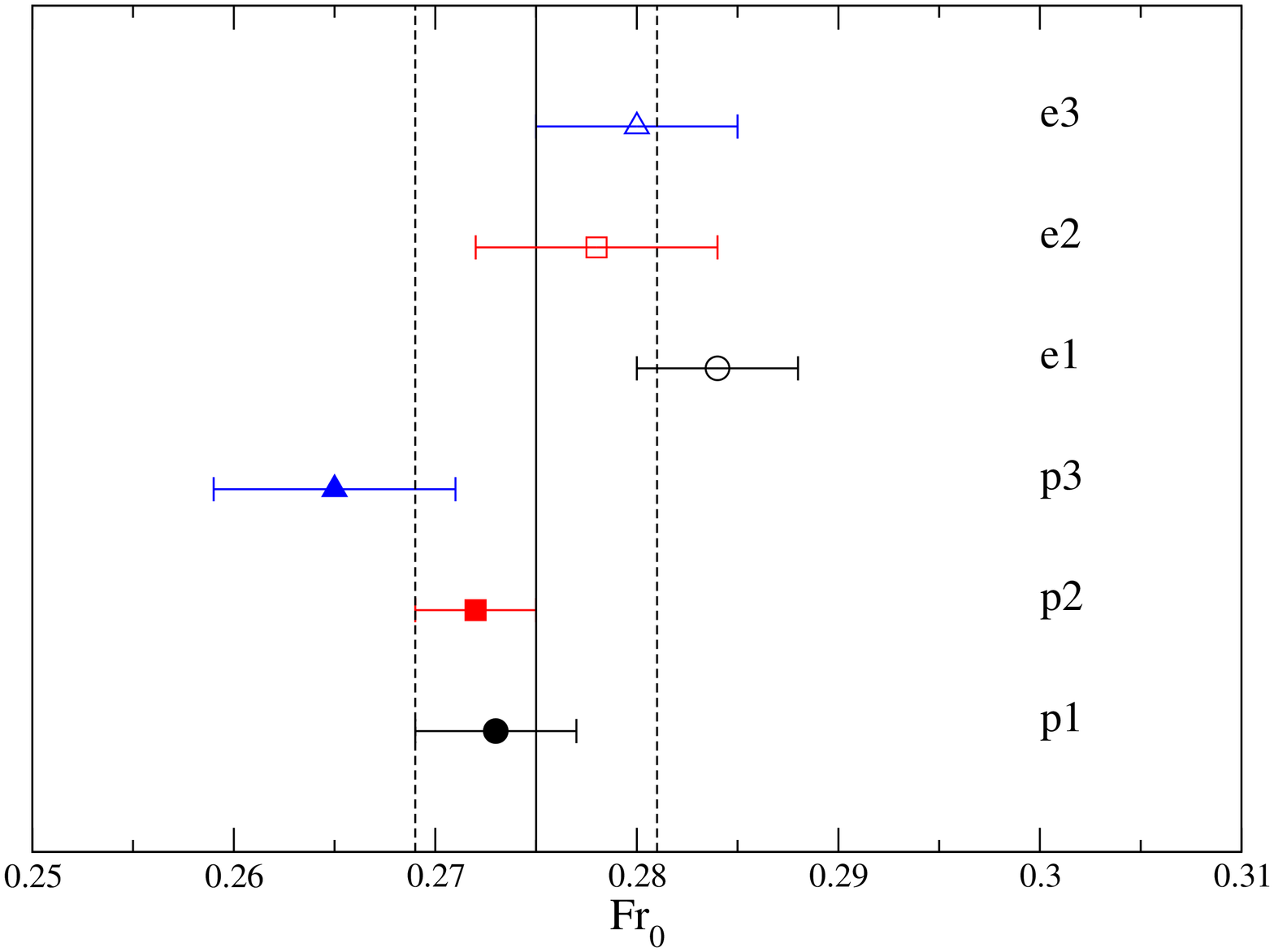}
\caption{Summary of the results for $F r_0$ obtained in the $\epsilon$- and $p$- regimes. The vertical band corresponds to the weighted average, with error corresponding to the largest uncertainty of the individual measurements.}\label{fig:fr0_tot}
\end{center}
\end{figure}

After analysing the data in the two regimes, we can now compare the results obtained for the leading-order LECs $\Sigma$ and $F$.\\
We first compare the results for $F r_0$ for the different lattices at our disposal; the data
from the $\epsilon$-regime, Eq.~(\ref{Fres_eps1}), and from the $p$-regime, Eq.~(\ref{Fr0_p1}),  are summarised in Fig.~\ref{fig:fr0_tot}. The agreement is rather good, within one-two standard deviations. The vertical line represents the weighted average
\begin{equation}
Fr_0=0.275(6),
\end{equation}
where the error is taken as the largest uncertainty associated to the individual measurements. We choose this rather conservative error estimate since we know from our analysis that systematic effects are below the individual statistical uncertainties.
Using the phenomenological value $r_0=0.5$ fm, one obtains $F=108.6(2.4)$ MeV.

\begin{figure}
\begin{center}
\includegraphics[width=8cm,angle=-90]{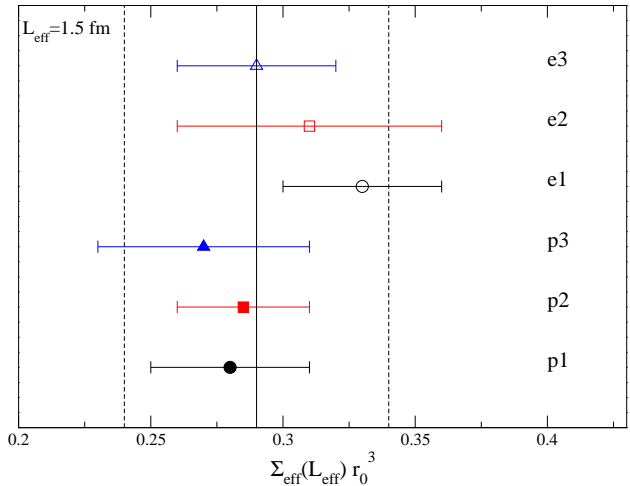}
\caption{Summary of the results for $\Sigma_{\rm eff}(L_{\rm eff}=1.5\;{\rm fm}) r_0^3$ obtained in the $\epsilon$- and $p$- regimes. The vertical band corresponds to the weighted average, with error corresponding to the largest uncertainty of the individual measurements.}\label{fig:sigma_tot}
\end{center}
\end{figure}

The results obtained for the quark condensate from $M_P^2/(2m)$ can be compared directly with the ones of a previous work \cite{Giusti:2007cn}, where the condensate has been computed in the $\epsilon$-regime from a finite-size scaling study.
The same parameters as for the lattices e1, e2, e3 have been adopted. 
The final results quoted in  \cite{Giusti:2007cn} for the renormalisation group invariant condensate at the scale
$L_{\rm eff}=1.5 $ fm are 
\begin{eqnarray} 
\Sigma_{\rm eff}(L_{\rm eff}=1.5\;{\rm fm})r_0^3 &=&  0.33(3),\;\;\; {\rm (e1)}\nonumber\\
\Sigma_{\rm eff}(L_{\rm eff}=1.5\;{\rm fm})r_0^3 &=&  0.31(5),\;\;\; {\rm (e2)}\label{cond_eps}\\
\Sigma_{\rm eff}(L_{\rm eff}=1.5\;{\rm fm})r_0^3 &=&  0.29(3).\;\;\; {\rm (e3)}\nonumber
\end{eqnarray}
In that analysis it was pointed out that, within the statistical uncertainty, no NLO volume-dependence is observed in $\Sigma_{\rm eff}$, as well as no lattice artifacts. We adopted  Eq.~(\ref{sigmaeff}) in order to express all results at the scale 1.5 fm.\\
Using the formula in Eq.~(\ref{sigma_rel}), we can convert the p-regime results in Eq.~(\ref{sigma_preg}) in $\Sigma_{\rm eff}(L_{\rm eff})$:
for the case $1/\mu=L_{\rm eff}=1.5$ fm we get
\begin{eqnarray} 
\Sigma_{\rm eff}(L_{\rm eff}=1.5\;{\rm fm})r_0^3 &=&  0.28(3),\;\;\;\;\; {\rm (p1)}\nonumber\\
\Sigma_{\rm eff}(L_{\rm eff}=1.5\;{\rm fm})r_0^3 &=&  0.285(25),\;{\rm (p2)}\label{cond_p}\\
\Sigma_{\rm eff}(L_{\rm eff}=1.5\;{\rm fm})r_0^3 &=&  0.27(4).\;\;\;\;\; {\rm (p3)}\nonumber
\end{eqnarray}
All these results are in good agreement, and are summarised in Fig.~\ref{fig:sigma_tot}. 
They can be converted into the $\overline{\rm MS}$- scheme at 2 GeV by using the relation \cite{Garden:1999fg}
$\overline{m}_{\overline{\rm MS}}(2\;{\rm GeV})/M=0.72076$. 
A weighted average gives
\begin{equation}
\Sigma_{\rm eff}^{\overline{\rm MS}}(2\;{\rm GeV})  =  (292\pm 17\;{\rm MeV})^3\;\;(L_{\rm eff}=1.5\;{\rm fm}),
\end{equation}
which lies in the same range of previous quenched computations \cite{Giusti:1998wy,Blum:2000kn,Hernandez:1999cu,Giusti:2001pk,DeGrand:2001ie,Hasenfratz:2002rp,Giusti:2003gf,Wennekers:2005wa,Bietenholz:2006fj}. Also in this case, the error associated to the average is the largest uncertainty coming from the single values.
     
\section{Conclusions}
We have computed the left-left current correlator in quenched lattice QCD, adopting
two different lattice spacings and two different physical volumes, with
quark masses in the $p$- and in the $\epsilon$-regime. We have compared our results with the expectations of the quenched chiral effective theory at NLO. \\
In the $\epsilon$-regime we have tested the LO finite-size scaling, which is well reproduced by the numerical data.
At NLO the effective theory predicts a time and topology dependence of the 
correlator: we have studied 
both effects and found good qualitative, but a rather poor 
quantitative, agreement with NLO predictions.
We should however point out that 
the statistical uncertainty is not good enough to measure the NLO 
effects with good precision.
Our results are compatible with the fact that higher order chiral corrections could be sizeable at these 
volumes.  We expect these corrections to
be more  important for larger $|\nu|$ and indeed the matching seems to
work worse there.\\
In the $p$-regime, we have extracted the pseudoscalar mass and decay constant from the two-point function. We have checked that the pseudoscalar decay constant at finite quark mass $F_P$ is volume-independent, as predicted by the effective theory, and from a chiral extrapolation we have extracted the low-energy constants $F$ and $\alpha_5$. 
We have studied the finite-volume effects on the pseudoscalar mass by 
computing ratios at different volumes.
We  have observed that after correcting our data for finite-volume effects, the behaviour of the quantity $M_P^2/(2m)$ as a function of the quark mass $m$ is well compatible with a constant one, with no signs of higher order terms, in the range of masses that we have considered. From $M_P^2/(2m)$ we have extracted the quark-condensate with $10\%$ precision. Finally we have compared the results obtained for $F$ from the $p$- and $\epsilon$-regime, finding a good agreement. Moreover, the results obtained for the condensate in the $p$-regime agree with the ones obtained in the $\epsilon$-regime from a previous study based on finite-size scaling.\\
This general agreement on the leading-order LECs is a non-trivial test that quenched QCD is well reproduced by quenched chiral effective theory at LO. On the other hand, a higher statistical accuracy would be needed to test the matching at NLO.

\section{Acknowledgments}
Our simulations were performed on PC clusters at CILEA, at DESY Hamburg, at the Universities of Rome ``La Sapienza'' and Valencia-IFIC, as well on the IBM MareNostrum at the Barcelona Supercomputing Center, and the IBM Regatta (JUMP) at FZ J\"ulich.
We thankfully acknowledge the computer resources
and technical support provided by all these institutions and by the University of Milano-Bicocca.\\
S.~N. is supported by Marie Curie Fellowship MEIF-CT-2006-025673.\\
S.~N. and P.~H. aknowledge partial support by Spanish Ministry for Education and Science under grant FPA2004-
00996 and Generalitat Valenciana under grant GVACOMP2007-156.\\
C.~P. acknowledges support from the Ram\'on y Cajal Programme, as well as by the
research grant FPA2006-05807 and the Consolider-Ingenio 2010 Programme CPAN
(CSD2007-00042) of the Spanish Ministry for Education and Science.\\
L.~G., P.~H., S.~N. and H.~W. are partially supported by EC Sixth 
Framework Program under the contract MRTN-CT-2006-035482 (FLAVIAnet).

\appendix
\section{Numerical results}\label{app_num}
In this appendix we report the results of the NLO fit according to Eq.~(\ref{eq:fit1})
of the left-left correlators computed in the $\epsilon$-regime.
\begin{table}
\begin{center}
\begin{tabular}{ |c |c| c| c|}
\hline
$am$ &   $|\nu|$  & $B_1$   &   $B_2$ \\  
\hline
0.001 &    0   &   0.103(22)  &   1.3(4)\\
      &    1   &   0.070(3)   &   1.43(18) \\
      &    2   &   0.0711(16) &   1.73(18) \\
      &    3   &   0.0703(16) &   2.72(22) \\
      &    4   &   0.0736(19) &   3.1(3)  \\
      &    5   &   0.0741(20) &   3.6(3)\\
\hline
0.003 &    0   &   0.083(12)  &   1.1(4) \\
      &    1   &   0.0697(25)   &   1.44(18) \\
      &    2   &   0.0711(16) &   1.73(18) \\
      &    3   &   0.0703(16) &   2.72(22)\\
      &    4   &   0.0737(19) &   3.1(3)  \\
      &    5   &   0.0741(20) &   3.6(3)\\ 
\hline
0.008 &    0   &   0.073(5)   &   1.2(3) \\
      &    1   &   0.0701(19)   &   1.53(17)\\
      &    2   &   0.0711(15) &   1.80(18)\\
      &    3   &   0.0704(16) &   2.76(22) \\
      &    4   &   0.0737(19) &   3.2(3) \\
      &    5   &   0.0741(20) &   3.7(3)\\
\hline
0.012 &    0   &   0.073(3)   &   1.3(3) \\
      &    1   &   0.0702(17)   &   1.64(17) \\
      &    2   &   0.0712(14) &   1.88(17) \\
      &    3   &   0.0706(15) &   2.81(22) \\ 
      &    4   &   0.0738(19) &   3.2(3) \\
      &    5   &   0.0742(20) &   3.7(3)\\
\hline
0.016 &    0   &  0.073(3)    &   1.52(24) \\
      &    1   &  0.0703(15)  &   1.78(16) \\
      &    2   &  0.0714(13)  &   1.98(17) \\
      &    3   &  0.0708(15)  &   2.87(22) \\
      &    4   &  0.0738(19)  &   3.3(3) \\ 
      &    5   &   0.0742(20) &   3.7(3)\\
\hline
\end{tabular}
\caption{Results for the fit in eq. \protect\ref{eq:fit1} for the lattice e1, in the time range $t/a=5-7$}\label{e1_res}
\end{center}
\end{table}
\begin{table}
\begin{center}
\begin{tabular}{|c| c| c| c|}
\hline
$am$ &   $|\nu|$  & $B_1$   &   $B_2$ \\
\hline
0.000316 & 0    &  0.08(3)        &    0.1(1.4)      \\
         & 1     &   0.076(9)       &   1.5(6)       \\
         & 2    &   0.068(3)       &  2.1(6)         \\
         & 3   &   0.0687(18)      &   2.7(4)       \\
         & 4    &   0.0696(19)      &   4.0(6)       \\  
         & 5   &  0.0706(16)      &   3.9(5)\\
         & 6   &  0.0683(15)      &  3.9(5)\\
         & 7   & 0.0675(20)  &     4.5(6)\\
\hline
0.000949  & 0    &  0.073(20)        &    0.3(1.1)      \\
         & 1     &   0.075(9)       &   1.5(6)       \\
         & 2    &   0.068(3)       &  2.1(6)         \\
         & 3   &   0.0687(18)      &   2.7(4)       \\
         & 4    &    0.0696(19)      &   4.0(6)       \\ 
         & 5    &   0.0706(16)      &   3.9(5)\\
         & 6    &  0.0683(15)      &  3.9(5)\\
         & 7    & 0.0675(20)  &     4.5(6)\\
\hline
0.00253 & 0   &  0.067(7)        &    0.4(8)      \\
        & 1     &   0.071(6)       &   1.7(5)       \\
         & 2    &   0.068(3)       &  2.2(5)         \\
         & 3   &   0.0687(17)      &   2.7(4)       \\
         & 4    &   0.0696(19)      &   4.0(6)       \\    
        & 5    &   0.0706(16)      &   3.9(5)\\
         & 6    &  0.0683(15)      &  3.9(5)\\
         & 7    & 0.0675(20)  &     4.5(6)\\
\hline
0.00380 & 0  &  0.067(5)        &    0.7(7)      \\
         & 1     &   0.070(4)       &   1.9(5)       \\
         & 2    &   0.069(3)       &  2.3(5)         \\
         & 3   &   0.0686(17)      &   2.8(4)       \\
         & 4    &   0.0696(18)      &   4.1(6)       \\    
       & 5    &   0.0707(16)      &   3.9(5)\\
         & 6    &  0.0680(15)      &  3.9(5)\\
         & 7    & 0.0676(20)  &     4.5(6)\\
\hline
0.00506 & 0 & 0.068(4)        &    1.0(6)      \\
         & 1     &   0.069(3)       &   2.0(5)       \\
         & 2    &   0.0690(25)       &  2.5(5)         \\
         & 3   &   0.0686(16)      &   2.9(4)       \\
         & 4    &   0.0696(18)      &   4.1(6)       \\      
         & 5    &   0.0707(15)      &   4.0(5)\\
         & 6    &  0.0681(15)      &  3.9(5)\\
         & 7    & 0.0676(20)  &     4.6(6)\\
\hline
\end{tabular}
\caption{Results for the fit in eq. \protect\ref{eq:fit1} for the lattice e2, in the time range $t/a=5-11$}\label{e2_res}
\end{center}
\end{table}

\begin{table}
\begin{center}
\begin{tabular}{|c |c |c| c|}
\hline
$am$ &   $|\nu|$  & $B_1$   &   $B_2$ \\
\hline
0.000612 &    0   &   0.064(14)  &   0.5(5)\\
      &    1   &   0.0516(20)    &   1.00(18) \\
      &    2   &   0.0516(14)  &     1.73(18) \\
      &    3   &   0.0526(12)  &   2.47(23) \\
      &    4   &   0.0524(16)   &   3.2(3)  \\
      &    5   &   0.0513(18)  &   3.1(4)\\
\hline
0.00184 &    0   &   0.058(9)  &   0.5(4) \\
        &    1   &   0.0516(20)   &   1.02(18) \\
        &    2   &   0.0517(14) &     1.74(18) \\
        &    3   &   0.0526(12) &    2.47(23) \\
        &    4   &   0.0524(16) &   3.2(3)  \\
        &    5   &   0.0513(18) &  3.1(4)\\
\hline
0.00490 &    0   &   0.053(4)   &   0.6(3) \\
        &    1   &   0.0520(17)   &   1.13(17) \\
        &    2   &   0.0518(13) &     1.80(18) \\
        &    3   &   0.0527(12) &   2.51(23) \\
        &    4   &   0.0524(15) &   3.2(3)  \\
        &    5   &   0.0513(18) &   3.1(4)\\
\hline
0.00735 &    0   &   0.053(3)   &   0.9(3) \\
       &    1   &   0.0522(15)   &   1.26(16) \\
        &    2   &   0.0519(13) &     1.87(18) \\
        &    3   &   0.0527(12) &   2.56(23) \\
        &    4   &   0.0525(15) &   3.2(3)  \\
         &    5   &   0.0514(18) &   3.1(4)\\
\hline
0.00980 &    0   &  0.0526(25)    &   1.1(3) \\
        &    1   &   0.0523(13)   &   1.41(16) \\
        &    2   &   0.0520(12) &     1.97(18) \\
        &    3   &   0.0528(12) &   2.63(23) \\
        &    4   &   0.0525(15) &   3.3(3)  \\
        &    5   &   0.0515(18) &   3.2(4)  \\
\hline
\end{tabular}
\caption{Results for the fit in eq. \protect\ref{eq:fit1} for the lattice e3, in the time range $t/a=6-10$}\label{e3_res}
\end{center}
\end{table}

\end{document}